%% file: main.tex
\documentclass[compsoc, conference, letterpaper, 10pt, times]{IEEEtran}

\input{setup/packages}

\input{setup/macros}

\IEEEoverridecommandlockouts
\IEEEpubid{\makebox[\columnwidth]{} \hspace{\columnsep}\makebox[\columnwidth]{ }}

\begin{document}
\title{A first look at browser-based cryptojacking}

\author{

\IEEEauthorblockN{Shayan Eskandari\IEEEauthorrefmark{1}, Andreas Leoutsarakos\IEEEauthorrefmark{1}, Troy Mursch\IEEEauthorrefmark{2}, Jeremy Clark\IEEEauthorrefmark{1}}

\IEEEauthorblockA{\IEEEauthorrefmark{1}Concordia University, \IEEEauthorrefmark{2}Bad Packets Report}

}

\maketitle
\IEEEpubidadjcol

\input{sections/abstract}

\IEEEpeerreviewmaketitle


\input{sections/intro}
\input{sections/body}


\bibliographystyle{abbrv}
\footnotesize
\bibliography{bib/references.bib}
\normalsize


\end{document}

%% file: setup/packages.tex
\usepackage{mdwmath}
\usepackage{mdwtab}
\usepackage{eqparbox}

\usepackage{fixltx2e}
\usepackage{stfloats}

\usepackage{bibspacing}
\usepackage[pdftex]{graphicx}
\graphicspath{{figures/}}
\DeclareGraphicsExtensions{.jpg,.png}

\usepackage[caption=false,font=footnotesize]{subfig}

\usepackage{amsmath}
\usepackage{bbm}
\usepackage{stmaryrd}

\usepackage{array}
\usepackage{color}
\usepackage[hyphens]{url}
\usepackage[pdftitle=Title,pdfauthor=Anonymous]{hyperref}
\usepackage{xcolor}

\usepackage{listings}

\lstdefinestyle{interfaces}{
  float=t,
  floatplacement=t}


%% file: setup/macros.tex
\usepackage{xspace}

\newcommand{\ie}{\textit{i.e.,}\xspace}
\newcommand{\eg}{\textit{e.g.,}\xspace}
\newcommand{\cf}{\textit{cf.}\xspace}



%% file: sections/abstract.tex

\begin{abstract}

In this paper, we examine the recent trend towards in-browser mining of cryptocurrencies; in particular, the mining of Monero through Coinhive and similar codebases. In this model, a user visiting a website will download a JavaScript code that executes client-side in her browser, mines a cryptocurrency---typically without her consent or knowledge---and pays out the seigniorage to the website. Websites may consciously employ this as an alternative or to supplement advertisement revenue, may offer premium content in exchange for mining, or may be unwittingly serving the code as a result of a breach (in which case the seigniorage is collected by the attacker). The cryptocurrency Monero is preferred seemingly for its unfriendliness to large-scale ASIC mining that would drive browser-based efforts out of the market, as well as for its purported privacy features. In this paper, we survey this landscape, conduct some measurements to establish its prevalence and profitability, outline an ethical framework for considering whether it should be classified as an attack or business opportunity, and make suggestions for the detection, mitigation and/or prevention of browser-based mining for non-consenting users. 

\end{abstract}


%% file: sections/intro.tex


\section{Introduction}

Bitcoin~\cite{nakamoto2008bitcoin} emerged almost a decade ago as an open source project, which mushroomed into a cryptocurrency sector collectively capitalized at over \$500 billion USD\footnote{Coinmarketcap - Global Charts - Accessed: 2017-12-14 \url{https://coinmarketcap.com/}}. Every day, people new to the concept of cryptocurrencies look for a quick and simple way to acquire some crypto-wealth. In the early days of Bitcoin, users on their personal computers could effortlessly acquire the currency through mining---a process Bitcoin uses to incentivize nodes to verify transactions as they are recoded in the blockchain. However, a second wave of mining technology saw users augmenting the CPU power of their computers with GPUs. Other groups of people deployed snippets of JavaScript code on websites that recruited their visitor's CPU power, often unknowingly, to mine for them as part of a bigger mining network (\ie a mining pool). However, both approaches quickly became infeasible as the computing power required to mine bitcoins grew exponentially to over 12 petahashes\footnote{Bitcoin hash rate - Accessed: 2017-11-20 \url{https://blockchain.info/charts/hash-rate}}. This was due to the emergence of application-specific integrated circuits (ASICs) and collective mining pools, which continue the third wave of mining to this day~\cite{narayanan2016}. 

\begin{figure}[t]
\centering
\includegraphics[width=0.9\linewidth]{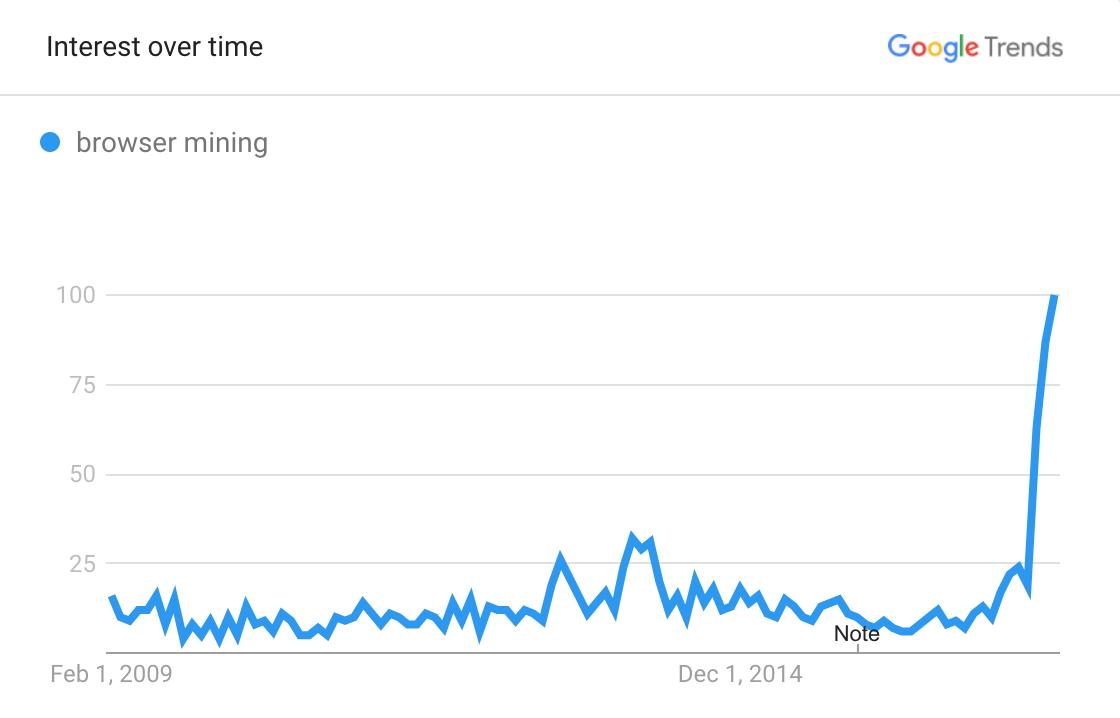}
\caption{Search interest for ``browser mining'' over time. Search interest seems to have piqued during price surges, which culminated with Bitcoin crossing \$1000 USD for the first time in December 2013. Soon after Bitcoin's first major crash searches consistently waned until a recent large spike, which is more than 4 times the lifetime average. The waning period before the recent surge could be attributed to the advent of ASIC usage for Bitcoin mining, and the surge is likely due to the revival of browser mining for non-Bitcoin currencies that have gained a sizeable market capitalization.\label{fig:interest}}
\end{figure}

As the years passed and a few key cryptocurrencies emerged as the market leaders, the concept of browser mining largely became forgotten. Today, the most common way for the average person to acquire cryptocurrencies is to purchase them. It came as a surprise to many when stories began to circulate on popular media outlets this year about websites mining cryptocurrencies through browsers again. Figure~\ref{fig:interest} shows how the searches for ``browser mining'' have changed since Bitcoin was launched. Websites like The Pirate Bay ~\cite{piratesbayhive} experimented with browser mining as a way to add a new revenue stream, while others like Showtime.com ~\cite{showtimehive} claimed they had the code injected after they were discovered. 

This paper tells the story behind the rejuvenation of browser-based mining. It is centred on \textit{cryptojacking} (also known as \textit{coinjacking} and \textit{drive-by mining}), a term coined to refer to the invisible use of a vulnerable user's computational resources to mine cyptocurrencies. Technically in-browser mining is a subset of cryptojacking, although most uses of the term apply to browser-based mining. In this case, mining happens within the client browser when the user visits the website. We have also seen the term cryptojacking applied to malware that mines cryptocurrencies, or in the situation where malware renders a machine as an unwitting participant in a botnet, and the botnet is rented for the purposes of mining crypcurrencies (\cf~\cite{huang2014botcoin}). The resource consumption of in-browser cryptojacking can noticeably degrade a computer's performance.

%
%
%
%
%
%

\section{Preliminaries and Related Work}

\subsection{Browser-based Mining}

\subsubsection{Early days}
The idea of in-browser mining started in the early days of Bitcoin. Bitcoin Plus\footnote{Bitcoin For the Uninitiated: Now, A Browser-Based Mining Client  May 19th, 2011 \url{https://www.themarysue.com/browser-based-bitcoin-mining/}} is one example of a discussion on replacing ads with Bitcoin browser miners\footnote{BitCoin browser mining as a replacement for ads \url{https://www.reddit.com/r/Bitcoin/comments/ieaew/bitcoin_browser_mining_as_a_replacement_for_ads/}}. It was also argued that browser-based mining provides greater scalability and decentralization as the barrier to entry is lowered to any unmodified computer with an internet connection. Soon after there was a rise in Bitcoin JavaScript miners such as JSMiner (2011)\footnote{A JavaScript Bitcoin miner \url{https://github.com/jwhitehorn/jsMiner}} and MineCrunch (2014)\footnote{MineCrunch, web(JS) miner with integration feature \url{https://cryptocurrencytalk.com/topic/24618-minecrunch-web-js-miner-with-integration-feature/}}. MineCrunch's visibility was increased by campaigns and the active online presence of its developers. Based on the developer claims, MineCrunch was well optimized for Javascript, but still worked 1.5x slower than native applications for CPU mining (\eg CPUMiner\footnote{CPU miner for Litecoin and Bitcoin - \url{https://github.com/pooler/cpuminer}}). Although CPU mining became uncompetitive with GPU and ASIC-based mining, it remained a sandbox for botnet admins to experiment with the thousands of CPUs at their disposal. Botnet mining has been studied in the literature~\cite{huang2014botcoin,wyke2012zeroaccess}, as well as covert mining within enterprises and cloud environments~\cite{MiningonSOeDime2017}.

In addition to unprofitability, browser-based mining faced legal challenges. In May 2015, the New Jersey Attorney General's office reached a settlement with the developers of ``Tidbit``, a browser-based Bitcoin miner. Terms of the settlement included ceasing operations of Tidbit. Then acting Attorney General John J. Hoffman stated ``No website should tap into a person's computer processing power without clearly notifying the person and giving them the chance to opt out.``~\cite{njcourtbitcoinjsminer}.

\subsubsection{From one CPU to ASICS and mining pools}

The first Bitcoin block mined on a GPU happened on July 18th, 2010 by a user named ArtForz~\cite{bitcoinhistory}, by using a private mining code that he developed himself. It was not until mid-2011 that others started implementing and releasing open source GPU-based mining tools. These tools greatly increased mining efficiency due to the hashing power of a GPU and the massive parallelizing possible with multiple GPUs (also known as mining rigs). The move from software to hardware followed shortly after. First, programmable FPGA chips resulting in custom-built circuits specifically for mining\footnote{Custom FPGA Board for Sale! (August 18, 2011) \url{https://bitcointalk.org/index.php?topic=37904.0}}. Then by mid-2012, companies started selling ASICs designed specifically for Bitcoin mining. After delay of about a year in delivering ASIC products, Bitcoin mining started transitioning from GPUs to ASICs where it remains today. Consequently, the hashing power of the Bitcoin network increased and the mining difficulty followed. To illustrate the change, consider a desktop PC CPU mining at 10 MH/s: on expectation, it will take 425 years before mining a single block~\cite{huang2014botcoin}. 

In parallel to the evolving technology, collective action emerged through the use of mining pools. A mining pool is a collective of individual miners. Participants receive a slice of work for mining the current block on behalf of the pool. If a member of the pool mines the block, the block reward is split amongst the participants of the pool \textit{pro rata} according to their computational effort~\cite{rosenfeld2011analysis}. As an aside, a very elegant protocol for reporting `near-solutions' to the pool enables participants to prove, without trust, the level of effort they are contributing to the pool at all times. In general, a mining pools cannot amplify earnings, they only change their shape. An income stream from a pool is a steady trickle, while solo-mining results in sporadic dumps of income. The first Bitcoin block found on a mining pool was on December 16, 2010 that was a beta implementation of a pool operated by a user named \textit{slush}.

\subsection{Monero}

Launched in April 2014, Monero~\cite{monero} is a cryptocurrency alternative to Bitcoin. It purportedly offers increased privacy by obfuscating the participants in a transaction, as well as the amounts. This is in contrast to more popular cryptocurrencies like Bitcoin and Ethereum, where a pseudonymous-but-complete transaction graph can be constructed from the public blockchain. Recent research has shown Monero's obfuscation techniques are less effective than originally claimed~\cite{MMLN17,kumar2017traceability}. Since regulation on exchanging between cryptocurrencies is lighter than exchanging cryptocurrencies for fiat money, and such services are not geographically bound, obtaining Monero for Bitcoin and vice versa is efficient and enables Monero to be used as a short-term medium of exchange for Bitcoin holders. This approach (and Monero's acceptance) is particularly popular on so-called dark web markets; markets that do not ban illicit goods and services.

A second characteristic that distinguishes Monero from Bitcoin is in the mining algorithm it uses. Monero still employs proof-of-work, specifically an algorithm called CryptoNight~\cite{cryptoknight}. However the computational puzzle is designed to be \textit{memory-hard}: it requires the storage of a large set of bytes and then requires frequent reads and writes from this memory. Such puzzles are optimized for CPUs with low-latency memory-on-chip, and not as well suited for circuits like FPGAs and ASICs. CrypoNight requires approximately 2MB per instance, which fits in the L3 cache of modern processors. Over the course of the next few years, these L3 cache sizes should become mainstream and allow more CPUs, and thus users, to participate in Monero's ecosystem. It has also been shown that ASICs cannot handle more than 1MB of internal memory, which is less than the size of memory required to calculate a new block. GPUs are also at a disadvantage since GDDR5 memory, which are used in modern GPUs and considered one of the fastest types of memory, is notably slower than L3 cache~\cite{van2013cryptonote}.  

%
%
%
%
%
%

\subsection{Coinhive}

\input{sections/timeline}

Monero built on its early success and continued to gain in popularity over the years, which caught the attention of some developers who decided to revisit the idea of browser mining. See Table~\ref{tab:timeline} for a timeline of events.  One of the earliest efforts appeared in September 2017 and was called Coinhive~\cite{coinhive}. Soon after, a competitor named Crypto-Loot\footnote{Crypto-Loot - A web Browser Miner | Traffic Miner | CoinHive Alternative \url{https://crypto-loot.com/}} emerged. Both websites provided APIs\footnote{Application Programming Interface} to developers for implementing browser mining on their websites that used their visitors' CPU resources to mine Monero. A portion of mined Monero would go back to the API developer, and the rest would be kept by the website. Not long after their early success, several copycats appeared such as Coin-Have and PPoi~\cite{coinhivecopycats} to take part in the reborn practice. It even inspired a new coin specifically designed for browser mining named JSECoin,\footnote{JSEcoin`s Website Cryptocurrency Mining \url{https://jsecoin.com/}} which has yet to find an audience. These developments took place over the course of a few weeks, which signalled the renewed success of browser mining. However, Coinhive's approach as a legitimate group set it apart from its peers and established itself as the leader in the space. They also launched separate services such as proof-of-work CAPTCHAs and short-links, which could be used to prevent spam while mining Monero~\cite{coinhive}.

%
%
%
%
%
%

%% file: sections/timeline.tex

\newcommand\ytl[2]{
\parbox[b]{8em}{\hfill{\color{cyan}\bfseries\sffamily #1}~$\cdots$~}\makebox[0pt][c]{$\bullet$}\vrule\quad \parbox[c]{4.5cm}{\vspace{7pt}\color{red!40!black!80}\raggedright\sffamily #2.\\[7pt]}\\[-2pt]}

\begin{table}
\centering
\begin{minipage}[t]{.7\linewidth}
\color{gray}
\rule{\linewidth}{1pt}
\ytl{2014-04-18}{Monero Cryptocurrency released}
\ytl{2017-09-14}{Coinhive Miner launched}
\ytl{2017-09-17}{ThePirateBay caught using coinhive~\cite{gaurdianelectricity}}
\ytl{2017-09-21}{Adblockers started to block coinhive scripts}
\ytl{2017-09-24}{Showtime caught running coinhive~\cite{showtimehive}}
\ytl{2017-09-25}{Coinhive clones started to appear}
\ytl{2017-10-13}{PolitiFact website compromised~\cite{politifactcoinhive}}
\ytl{2017-10-16}{Coinhive launched authedmine - authorized mining}
\ytl{2017-11-23}{LiveHelpNow Hack incident~\cite{chatsupporthack}}
\ytl{2018-01-25}{Cryptojacking code found on Youtube ads~\cite{trendmicrocoinhive}}
\ytl{2018-02-11}{UK Information Commissioner's Office incident~\cite{theregisterukgovern}}
 
\rule{\linewidth}{1pt}%
\end{minipage}%
\bigskip
\caption{Timeline of Monero and in-browser mining reports\label{tab:timeline}}
\end{table}

%% file: sections/body.tex

%
%
%
%
%
%

\section{Threat Model}
In-browser mining is considered as an abuse unless user's consent is granted. The attack surface to abuse users' browsers through cryptojacking is broad, and there are multiple vectors where various entities can inject mining scripts in the website's codebase. We summarize those here. 

\subsection{Webmaster initiated} 

A website administrator can add a mining script to her webpage, with or without informing users. Website owners may do this to monetize their sites, especially when they have been blacklisted or blocked by standard advertising platforms. In one example, a researcher found Coinhive on a large Russian website offering child pornography to users~\cite{coinhiveonchildporn}. Revenue estimates, based on the website traffic data available, were roughly \$10,000 a month after converting the value of XMR mined to USD.

\subsection{Third-party services} 

Many websites serve active Javascript from third parties within their own webpages. This could be ads from an ad network, accessibility tools or tracking and analytics services. Third parties with these privileges can inject cryptojacking scripts into the sites that use them, either intentionally or as a result of a breach. The first two incidents, Coinhive was injected into the websites of Movistar\footnote{Movistar is a major telecommunications brand owned by Telefonica, operating in Spain and in many Hispanic American countries \url{https://www.movistar.com/}} and Globovision\footnote{Globovision is a 24-hour television news network in Venezuela and Latin America \url{http://globovision.com/}} using Google Tag Manager\footnote{Google Tag Manager is a tag management system created by Google to manage JavaScript and HTML tags used for tracking and analytics on websites}. Movistar stated that Coinhive was not put on their website by a hacker, but instead was due to \textit{``internal error''} while they were conducting ``production tests''. No statement was provided by Globovision on why the cryptojacking scripts appeared on their site on November 15, 2017~\cite{googletagcoinhive}. Another high-profile cryptojacking case involving a Google platform occurred in January 2018 when Trend Micro researchers found advertisements containing Coinhive miner script were shown to YouTube users in Japan, France, Taiwan, Italy, and Spain for nearly a week~\cite{trendmicrocoinhive}. Similar to Showtime, YouTube inherently has a high average visit duration as a video streaming site and thus is prime target for cryptojacking operations.

\subsection{Browser extensions} 

Cryptojacking was not limited to websites in 2017. The Chrome extension \textit{Archive Poster} remained on the Chrome Web Store for days while silently cryptojacking an unknown portion of their 100,000+ users. After multiple user reports, followed by multiple news media outlets covering the issue, the extension was removed~\cite{chromeextentioncoinhive}. Similar cryptojacking extensions has been identified on less popular Mozilla Firefox add-ons as well.

\subsection{Breaches} 

If an attacker is able to breach principle servers, websites, extensions, or the scripting services they use, they can inject cryptojacking scripts that will impact the site's users without the site's knowledge or consent. For example, a researcher found a malicious modification to webchat system LiveHelpNow's SDK; it resulted in unsolicited mining across nearly 1500 websites using their chat support service~\cite{chatsupporthack} such as retail store chains Crucial and Everlast websites. In another example, Coinhive was found on the political fact-checking website PolitiFact\footnote{PolitiFact: Fact-checking US politics \url{https://politifact.com/}} A compromised JavaScript library was found to be injecting the cryptojacking code. The malicious code remained on the site for at least four hours before it was removed~\cite{politifactcoinhive}. PolitiFact executive director stated, ``Hackers were able to install their script on the fact-checking website after discovering a misconfigured cloud-computing server''~\cite{politifactcoinhivewsj}. 

Another recent example of such incident is a breach in a website plugin called \textit{Browsealoud}\footnote{An accessibility tool to read the content aloud in multiple languages \url{https://www.texthelp.com/en-gb/products/browsealoud/}} led to injection of cryptojacking scripts in some United Kingdom governmental websites such as \textit{Information Commissioner`s Office, UK NHS services, Manchester City Council} and around 4200 other websites~\cite{theregisterukgovern}. Within the same month, cryptojacking script was seen on \textit{Tesla} and \textit{LA Times} websites through poorly secured cloud configuration~\cite{nakedsecurityLatimes}.

\subsection{Man-in-the-middle} 

A user's web traffic is often routed through intermediaries that may have plaintext access to content. For example, internet service providers or free public wireless routers can inject cryptojacking scripts into non-HTTPS traffic. Advertisement code injection has been seen in practice before~\cite{vergeadinjection} and there have been assertions of similar injections of browser mining scripts at certain Starbucks free Wi-Fi hotspots in Argentina\footnote{\url{https://twitter.com/imnoah/status/936948776119537665}}.

%
%
%
%
%
%

\section{Measurements}

\subsection{Prevalence of Coinhive and alternatives}

\begin{figure}[h]
\centering
\includegraphics[width=\linewidth]{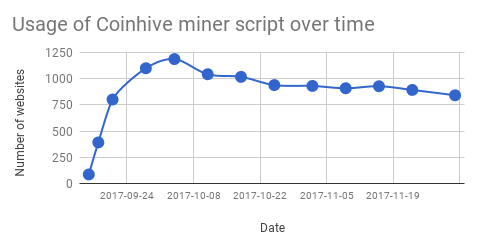}
\caption{The number of instances of the Coinhive miner scripts found using the query in Figure~\ref{lst:bigquery} in top one million websites over a three month period beginning with the release of Coinhive in September 2017.\label{fig:topmil}}
\end{figure}

\begin{figure*}[t]
\begin{tabular}{c}
\begin{lstlisting}[language=sql]
SELECT domain, tags, p80.http_www.get.headers.content_language, p80.http_www.get.headers.server, p80.http.get.headers.x_powered_by, p80.http.get.title, p80.http_www.get.body as wwwbody, p80.http.get.body as plainbody 
FROM censys-io.domain_public.20171123
WHERE STRPOS(p80.http.get.body, coinhive.min.js) > 0 or STRPOS(p80.http_www.get.body, coinhive.min.js) >0)
\end{lstlisting}
\end{tabular}
\caption{A BigQuery SQL query to find websites that embed the Coinhive script using a dataset of the top one million sites from censys.io. \label{lst:bigquery}}
\end{figure*}

\begin{figure}[t]
\centering
\includegraphics[width=0.9\linewidth]{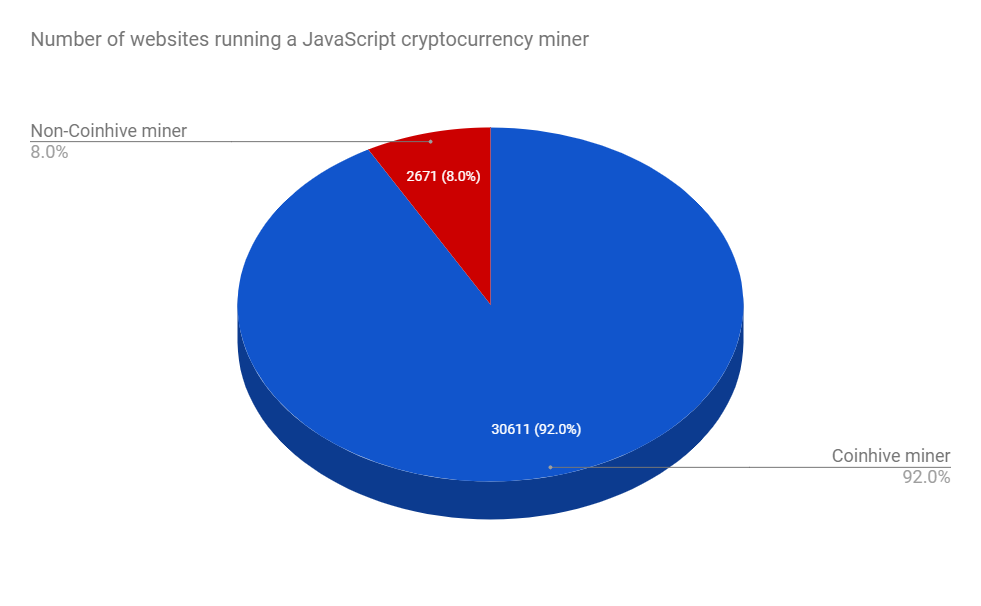}
\caption{Share of websites using a Javascript cryptocurrency miner, details in Table~\ref{tab:findcoinhive}  \label{fig:coinhivevscopycats}}
\end{figure}

Based on the fact that Coinhive is the dominant website offering in-browser mining  (see Figure~\ref{fig:coinhivevscopycats}), we first focus on measuring the prevalence of Coinhive scripts deployed on internet sites. We use the censys.io BigQuery dataset~\cite{censys15} for the top million sites indexed by Zmap\footnote{\url{https://zmap.io}}. We simply look for the \texttt{coinhive.min.js} script within the body of the website page. The query we use is in Figure~\ref{lst:bigquery} and the results over a two month period are provided in Figure~\ref{fig:topmil}. These findings are corroborated by another search engine, PublicWWW\footnote{Search Engine for Source Code \url{https://publicwww.com/}}, which indexes the source code of publicly available websites. Using PublicWWW's dataset, over 30,000 websites were found to have the \texttt{coinhive.min.js} library~\cite{badpacketspublicwww}. As seen from our data in Figure~\ref{fig:topmil}, the adoption of this script was substantial in the first days of its release. However, progress slowed down at the same time as ad-blockers and organizations started to block Coinhive's website. The initial purpose of this service, as claimed by Coinhive, was to replace ads and cover server costs for webmasters. As the service did not require that websites receive user consent before running the miner code, it started to be used maliciously in users` browsers. This type of usage resulted in Coinhive being included in some company's top-10 most wanted malware list~\cite{checkpoint}.

This type of measurement will become less accurate moving forward. Cryptojacking services are evolving to use obfuscated JavaScript and randomized URLs to evade detection\footnote{\url{https://twitter.com/bad_packets/status/940333744035999744}}. An example of these methods can be found in the cryptojacking service provider called Minr. In this case, the script is automatically obfuscated for users implementing the code. In addition, the domain names used by Minr frequently change to circumvent blocklists and anti-malware software.

\begin{figure}[t]
\centering
\includegraphics[width=0.9\linewidth]{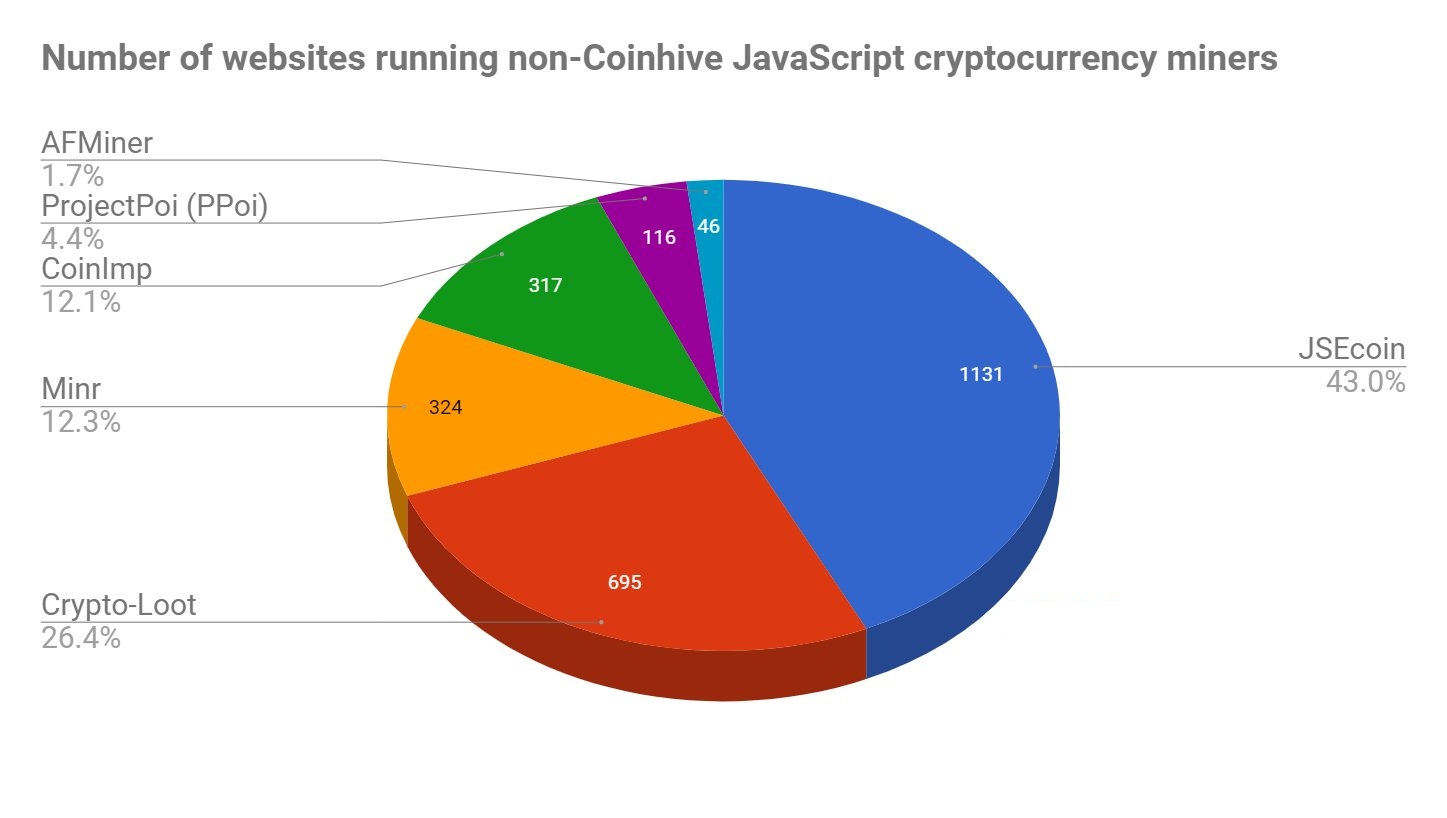}
\caption{Share of websites using a Coinhive alternative, details in Table~\ref{tab:findcoinhive} } \label{fig:copycat}
\end{figure}

\begin{figure}[t]
\centering
\includegraphics[width=0.9\linewidth]{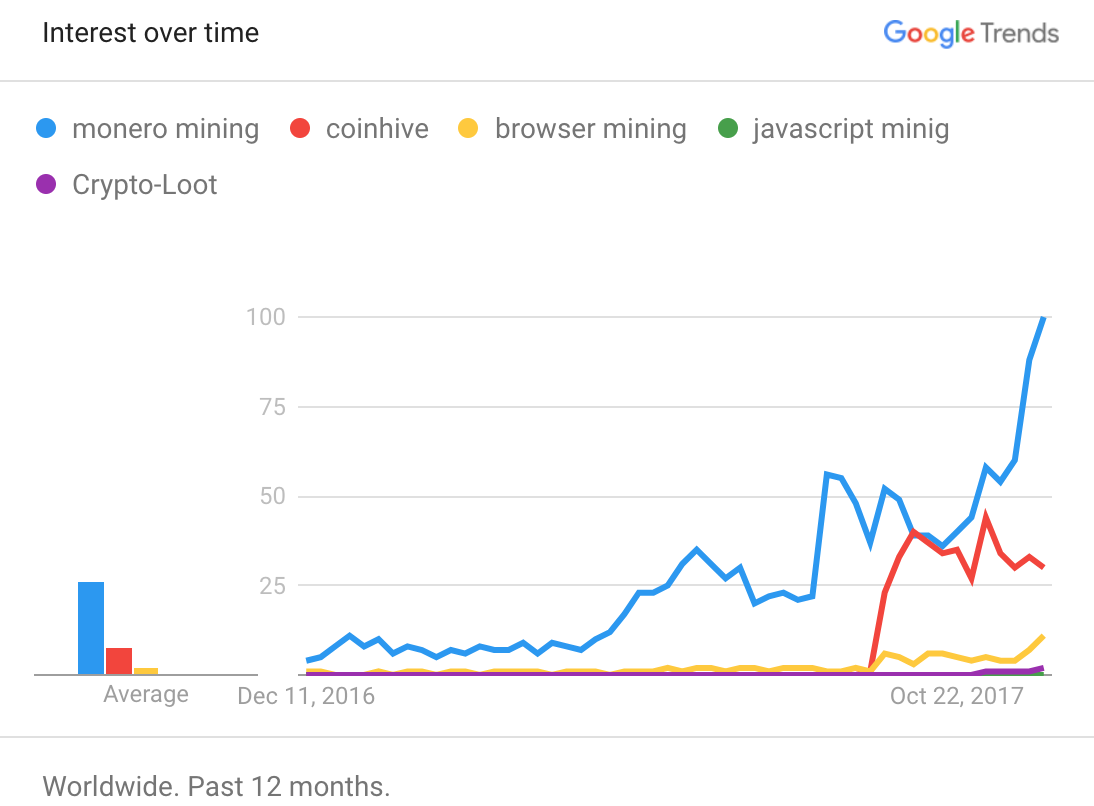}
\caption{Google Trend over last 12 months: there has been more interest in Coinhive than the broader, related search term ``Browser mining''. Comparing to other services offering Monero browser mining API, Coinhive had the advantage of being the first to offer the service. \label{fig:trend}}
\end{figure}

\begin{figure}[h]
\centering
\includegraphics[width=0.9\linewidth]{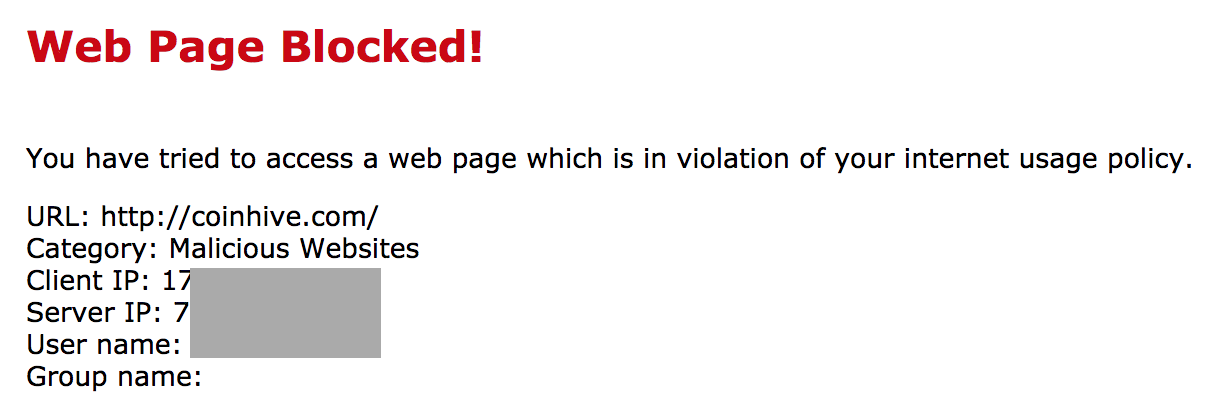}
\caption{Concordia university has categorized the coinhive.com website as malicious and has blocked it.\label{fig:concordia}}
\end{figure}

Coinhive has begun to be blocked by enterprises. One example is shown in Figure~\ref{fig:concordia}. This blocking seems to have sent Coinhive operators to lesser known alternatives with the same or similar functionality. We used the same methodology on PublicWWW dataset to find the usage of Coinhive and its alternatives on the internet. Table~\ref{tab:findcoinhive} shows the keywords used to identify these services. The result can be found on Figure~\ref{fig:coinhivevscopycats} and Figure~\ref{fig:copycat}.

\begin{table}[h]
\centering
\begin{tabular}{|cc|c|}
\hline
\textbf{Website} & \textbf{Results} & \textbf{Query Parameter}  \\ \hline
Coinhive & 30611 & `coinhive.min.js`   \\  \hline
JSEcoin & 1131 & `load.jsecoin.com`   \\  \hline
Crypto-Loot & 695 & `CryptoLoot.Anonymous`  \\  \hline
Minr & 324 & `minr.pw`,`st.kjli.fi`, \\
~  & ~ &  `abc.pema.cl`,`metrika.ron.si`, \\  
~  & ~ &  `cdn.rove.cl`,`host.d-ns.ga`, \\  
~  & ~ &  `static.hk.rs`,`hallaert.online`, \\  
~  & ~ &  `cnt.statistic.date`,`cdn.static-cnt.bid` \\     \hline
CoinImp & 317 &`www.coinimp.com/scripts/min.js`,  \\ 
~  & ~ &  `www.hashing.win` \\     \hline
ProjectPoi (PPoi) & 116 & `projectpoi.min`  \\  \hline
AFMiner & 46 & `afminer.com/code/miner.php`	  \\  \hline
Papoto & 42 & `papoto.com/lib/papoto.js`  \\ \hline
\end{tabular}
\bigskip
\caption{Cryptojacking data was gathered by totalling the number of websites which had the following libraries in their source codes, indexed by PublicWWW by 12/24/2017. Figure~\ref{fig:coinhivevscopycats} and Figure~\ref{fig:copycat} are visualizations of these result.\label{tab:findcoinhive}}
\end{table}

\begin{figure}[h]
\centering
\includegraphics[width=0.9\linewidth]{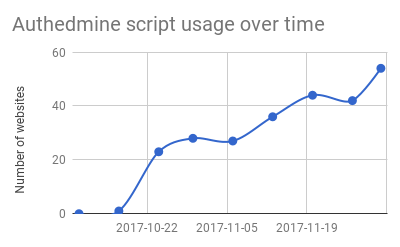}
	\caption{Usage of AuthedMine Miner scripts in top one million websites since its introduction} \label{fig:authmine}
\end{figure}

Coinhive has also reacted by focusing on adding methods to enforce asking for user consent and legitimizing the use of cryptojacking. It introduced another domain and service called \textit{Authedmine}, which requires user's consent to start mining in the browser. This service did not get the same attention as the original service, but it did inspire discussions regarding the ethics of such services, which is discussed in Section~\ref{sec:ethics}. Using the same methodology, censys.io was used to measure the prevalence of AuthedMine and show the results in Figure~\ref{fig:authmine}. 

\subsection{Client impact}

\begin{figure}[h]
\centering
\includegraphics[width=\linewidth]{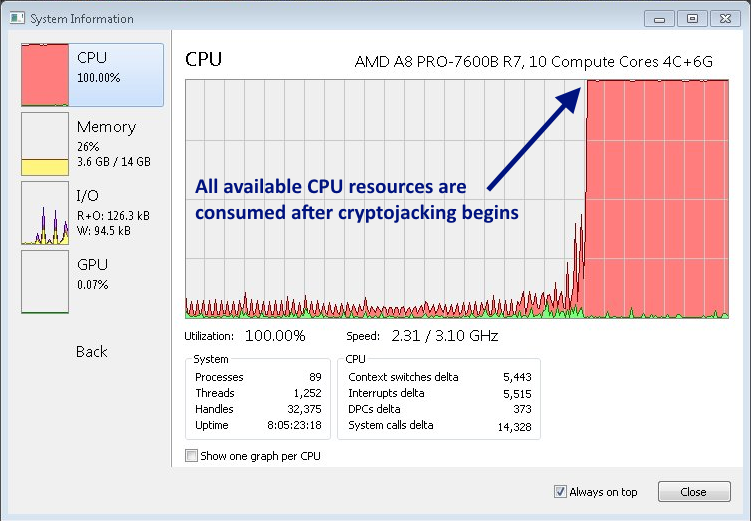}
	\caption{Comparison of CPU usage of browser without and with browser mining enabled.}\label{fig:cpu}
\end{figure}

Most cryptojacking scripts discovered were configured to use around 25\% of user's CPU, which can be justified as it will be under the threshold of attracting the user's attention, and it could be argued as fair-usage of their hardware. During the first few days, however, there were some reports of 100\% CPU usage while visiting websites containing these scripts~\cite{piratesbayblog}, which can be characterized as malicious. By default, the Coinhive JavaScript library will use all available CPU resources. The user implementing the script must include a throttle value to reduce the client-side CPU usage during mining operations. We show an example in Figure~\ref{fig:cpu}.

\subsection{Profitability}
\label{profitabilitexperiment}

\begin{figure}[t]
\centering
\includegraphics[width=\linewidth]{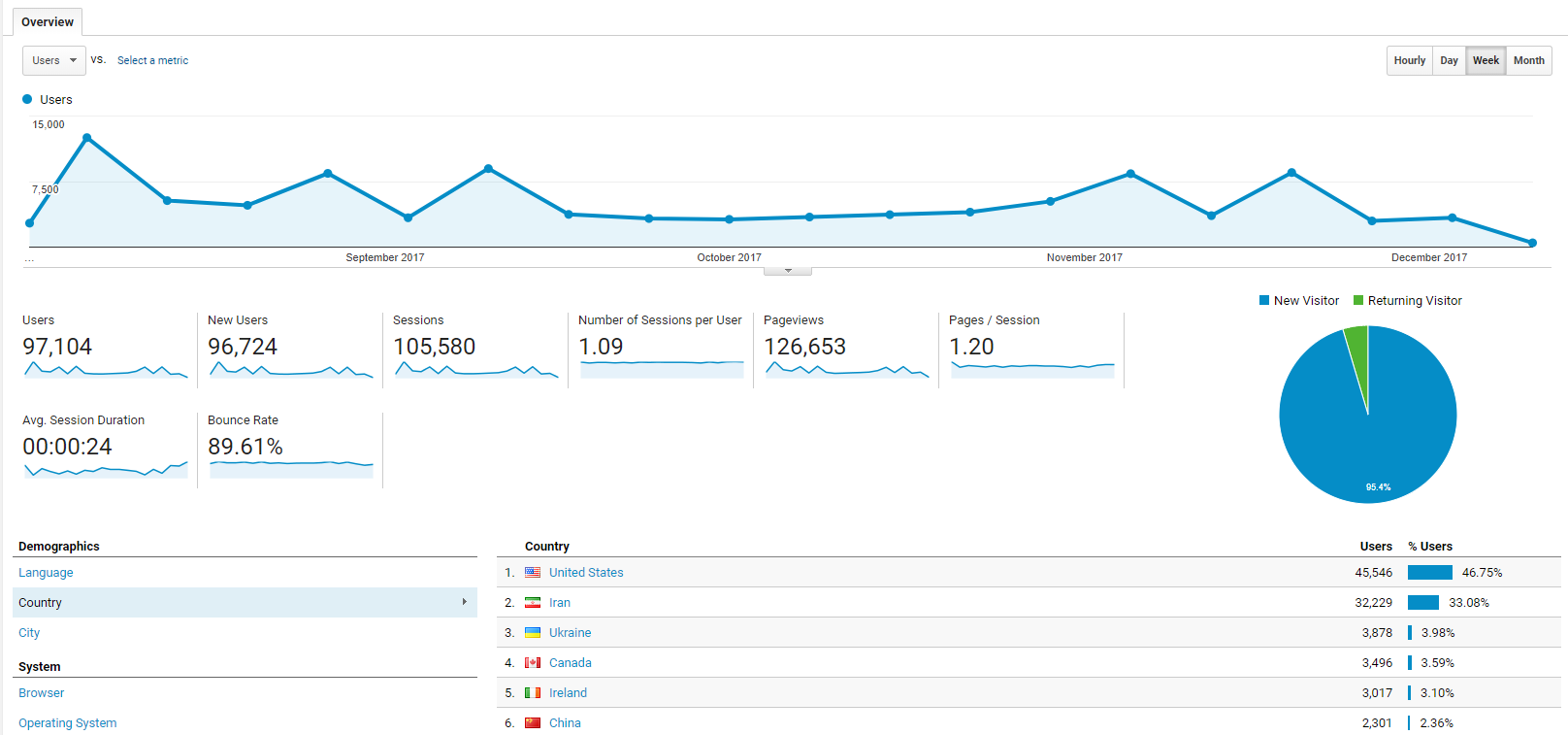}
\caption{Google Analytics dashboard showing the number of visitors to a domain parking service of 11\,000 domains.}\label{fig:domain2}
\end{figure}

\begin{figure}[t]
\centering
\includegraphics[width=\linewidth]{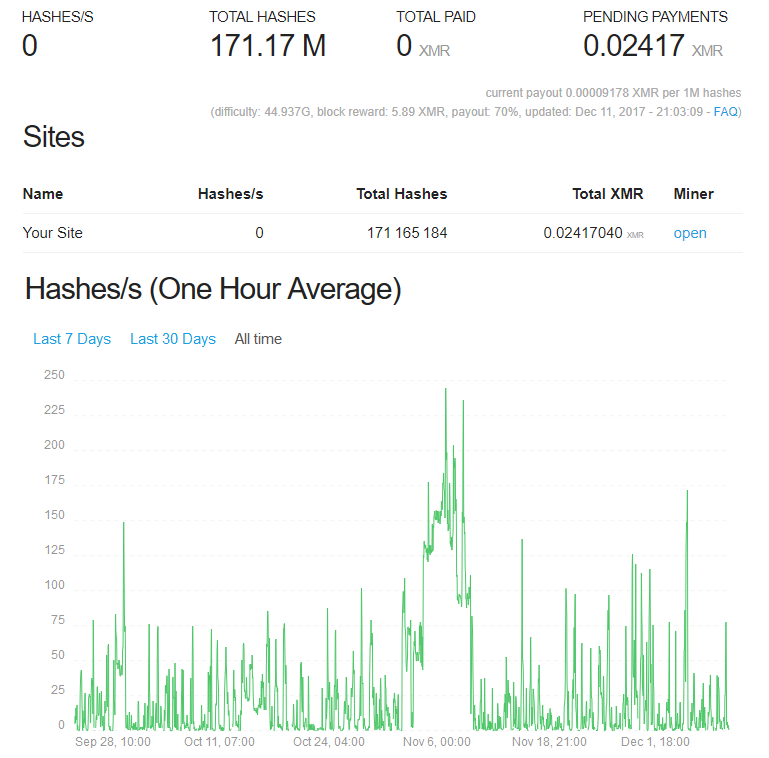}
\caption{Coinhive dashboard showing the earnings of a domain parking service that runs Coinhive on 11\,000 domains. Over the course of about 3 months, the operator earned 0.02417 XMR (currently \$7.69 USD).}\label{fig:domain1}
\end{figure}

Coinhive developers estimate a monthly revenue of about 0.3 XMR (about \$101 USD) for a website with 10-20 active miners~\cite{coinhive}. We sought to validate this estimation with a real world data set provided to us\footnote{In collaboration and with thanks to Faraz Fallahi \url{https://github.com/fffaraz}}. One of the biggest Coinhive campaign operators is a domain parking service. It runs Coinhive on over 11\,000 parked websites. While visits to parked domains are considerably shorter than an average website, the data spans a period of three months and gives some insight into the profitability of cryptojacking. During the experimental period of about 3 months, they accumulated 105\,580 user sessions for an average of 24 seconds per session. For the period examined, the revenue was 0.02417 XMR (Monero's currency) which at the time of writing is valued at \$7.69 USD. Further detail is provide in Figures~\ref{fig:domain2} and \ref{fig:domain1}. While an A/B test was not setup to determine how much traditional web advertising would have brought in, freely available web calculator tools suggest we might expect an order or two of magnitude greater for comparable traffic. 

\section{Mitigations}

We discuss the ethics of cryptojacking in the next section, but in the case of cryptojacking without user consent, it is seems natural to us to presuppose users want to be protected. Protection might take a few forms, which we outline here.

\subsection{Obtaining consent}

Cryptojacking tools might attempt to legitimize the practice by first obtaining user consent on a service provider level. An example of this is the Authedmine service from Coinhive discussed previously. Malicious sites might also opt for a service like Authedmine if it is whitelisted on its users` networks and then attempt to circumvent the consent process. For example, consent that requires a click from the user has been shown in some circumstances to be vulnerable to clickjacking attacks~\cite{rydstedt2010busting}.

While cryptojacking is nowhere near the prevalence of tracking cookies, eventually it might grow into a regulatory issue where governmental bodies could use legislative approaches to obtain consent, similar to the provisions many countries now use for cookies (including honouring the `do not track' HTTP header and obtaining click-based consent).


\subsection{Browser-level mitigation}

Browser developers have begun discussion of intervening in cryptojacking\footnote{`Please consider intervention for high cpu usage js' \url{https://bugs.chromium.org/p/chromium/issues/detail?id=766068}}. Potential mitigations include: throttling clientside scripting, warning users when clientside scripting consumes excessive resources, and blocking the sources of known cryptojacking scripts. Determining appropriate for thresholds for client-side processing that are high enough to allow legitimate applications and low enough to deter cryptojacking is an open research problem, as would be the wording of any notifications to the user that would lead the user to make an informed decision about allowing or not allowing resource consumption (\cf SSL/TLS warnings~\cite{SEAAC09,SHB11,Acer:2017:WWR:3133956.3134007}).
Browsers such as Opera, have taken a stance against cryptojacking scripts and blocked them via their ``NoCoin'' blacklist~\cite{operanocoin}. It is too early to determine the effectiveness of using a blacklist to block such activities.  

It is worth noting that some browsers might actually take the exact opposite approach and promote (consensual) in-browser mining, as it enables a form of monetizing websites independent of both (1) ad networks and the user tracking that accompanies the current ad model, and (2) users maintaining some form of credits or currencies for making micropayment to websites they use(\eg Brave Browser \footnote{\url{https://brave.com}}). Browser mining has been shown to not be as efficient as native mining applications today. Therefore, optimizations on how browsers pass system calls to the operating system can be made, or there can even be browsers designed specifically to support efficient browser mining.


\section{Discussion}
\label{sec:ethics}

While cryptojacking might be relatively new, it fits the pattern of various other technologies deployed on the web that raise ethical questions. In thinking about it, we distinguish a few cases: (1) the use of cryptojacking on a breached website, (2) the use of cryptojacking by the website owner with an attempt at obtaining user consent, and (3) the use of cryptojacking by the website owner without obtaining user consent. We would argue that (1) is clearly unethical; invariant to one's views on the ethics of hacking, we cannot see a justification for a breach that profits the adversary without any external benefits to anyone else.

The second case, cryptojacking after gaining user consent, is controversial primarily because it is unclear if users understand what they are consenting to, what they receive in return (some examples might include the elimination of ads, premium features, paywalled content, or higher definition video streams), and whether it is a fair exchange. To understand the zeitgeist, consider a recent poll conducted by Bleeping Computer that found: ``many users said they are OK with websites mining Monero in the background if they don't see ads anymore''~\cite{bleepingcomputerminers}. Coinhive released AuthedMine in recognition of the importance to many of user consent. ThePirateBay.org~\cite{bbcmintcrypto} ran cryptojacking scripts while users searched for torrent files without notice in their Privacy Policy, nor any visible warning on any part of the website that informed their users of this activity. This resulted in a backlash against the website, which responded with the following statement, ``Do you want ads or do you want to give away a few of your CPU cycles every time you visit the site?''~\cite{piratesbayblog}. While the admins admitted to their testing of browser mining, their notice came after it was discovered and they ultimately removed the code. In both auction-based and keyword-based online advertisement, the advertiser pays the advertisement publisher to distribute the advertisement and the advertisement publisher pays a portion of the revenues to the website owner whom the advertisement was shown on her website~\cite{king2007internet}. However with in-browser mining as a replacement monetization strategy, a more direct compensation is established with less intermediaries which could benefit users and sites alike. 

The potential harm to users of cryptojacking is higher energy bills, along with accelerated device degradation, slower system performance, and a poor web experience~\cite{httparchiveminingimpact,gaurdianelectricity}. While consent may be obtained from the user, it is unclear if the user's mental model of how they are paying can be made clear to them. On the other hand, the privacy disclosures users make in the traditional advertising model are also intangible; it is doubtful users understand what they are consenting to when they, for example, consent through a banner~\cite{eucookie} to the use of tracking cookies; and many websites waste computational resources without consequence through buggy scripting and unnecessary libraries. In short, the ethics are not clear-cut and should be debated. 

One webservice prone to cryptojacking is video streaming---the longer a user is engaged on a website, the more income can be earned through browser mining. Showtime.com~\cite{registershowtime} and UFC.com~\cite{registerufcmonero} are two popular streaming sites that were asserted by researchers to have deployed Coinhive. Showtime has declined to comment on how or why Coinhive was implemented on their website. Speculation has been raised that it was injected via a third-party analytic tool, New Relic, due to Coinhive being found inside the New Relic code block within showtime's website source code. However a New Relic representative denied these claims in a statement to The Register, ``It appears [Coinhive scripts] were added to the website by [Showtime's] developers.''~\cite{registershowtime}. In a statement released by the UFC, they denied the presence of the code stating, ``[they] did not find any reference to the mentioned Coinhive JavaScript [code]''\footnote{\url{https://twitter.com/bad_packets/status/928044219222048769}}.

The third case is the use of cryptojacking without user consent. Moor, in "What is Computer Ethics?"~\cite{moor1985computer} introduces the concept of an \textit{invisible factor} for invisible computer operations in society. Based on his definitions, we would classify cryptojacking that does not gain user consent as \textit{invisible abuse}: the intentional use of the invisible operations of a computer to engage in unethical conduct. Here the cryptojacker is earning money from unaware users that are being charged on their electricity bill. As discussed before, we already have court cases against such activities~\cite{njcourtbitcoinjsminer} and regulations for activities such as online user tracking~\cite{eucookie}, which indicates the need to start discussions and regulation on in-browser mining to fill in this policy vacuum as well. 


\section{Acknowledgements}

J. Clark thanks NSERC and FRQNT for partial funding of this research.